\documentclass[twocolumn]{aastex63}
\pdfoutput=1 
\usepackage{amsmath,amstext}
\usepackage[T1]{fontenc}
\usepackage{apjfonts}
\usepackage{hyperref}
\usepackage[all]{hypcap}
\usepackage{url}
\usepackage{xspace}
\usepackage{color}

\usepackage{savesym}
\savesymbol{tablenum}
\usepackage{siunitx}
\restoresymbol{SIX}{tablenum}

\usepackage{lineno}

\usepackage{booktabs}

\newcommand{\Spider}{\textsc{Spider}\xspace}

\newcommand{\Planck}{\textit{Planck}\xspace}

\shorttitle{Mode-Coupling Correction of CMB Power Spectra}
\shortauthors{\Spider Collaboration}

\begin{document}

\title{A Simulation-Based Method for Correcting Mode Coupling in CMB Angular Power Spectra}

\newcommand\ANL{High Energy Physics Division, Argonne National Laboratory, Argonne, IL, USA 60439}
\newcommand\CWRU{Physics Department, Case Western Reserve University, 10900 Euclid Ave, Rockefeller Building, Cleveland, OH 44106, USA}
\newcommand\Cardiff{School of Physics and Astronomy, Cardiff University, The Parade, Cardiff, CF24 3AA, UK}
\newcommand\UBC{Department of Physics and Astronomy, University of British Columbia, 6224 Agricultural Road,
Vancouver, BC V6T 1Z1, Canada}
\newcommand\Princeton{Department of Physics, Princeton University, Jadwin Hall, Princeton, NJ 08544, USA}
\newcommand\Caltech{Division of Physics, Mathematics and Astronomy, California Institute of Technology, MS 367-17, 1200 E. California Blvd., Pasadena, CA 91125, USA}
\newcommand\JPL{Jet Propulsion Laboratory, Pasadena, CA 91109, USA}
\newcommand\CITA{Canadian Institute for Theoretical Astrophysics, University of Toronto, 60 St.\@ George Street, Toronto, ON M5S 3H8, Canada}
\newcommand\ASU{School of Electrical, Computer, and Energy Engineering, Arizona State University, 650 E Tyler Mall, Tempe, AZ 85281, USA}
\newcommand\UKZN{School of Mathematics, Statistics and Computer Science, University of KwaZulu-Natal, Durban, South Africa}
\newcommand\NITP{National Institute for Theoretical Physics (NITheP), KwaZulu-Natal, South Africa}
\newcommand\Imperial{Blackett Laboratory, Imperial College London, SW7 2AZ, London, UK}
\newcommand\Stockholm{The Oskar Klein Centre for Cosmoparticle Physics, Department of Physics, Stockholm University, AlbaNova, SE-106 91 Stockholm, Sweden}
\newcommand\Oslo{Institute of Theoretical Astrophysics, University of Oslo, P.O. Box 1029 Blindern, NO-0315 Oslo, Norway}
\newcommand\TorontoDunlap{Dunlap Institute for Astronomy and Astrophysics, University of Toronto, 50 St.\@ George Street, Toronto, ON M5S 3H4 Canada}
\newcommand\Toronto{Department of Astronomy and Astrophysics, University of Toronto, 50 St.\@ George Street, Toronto, ON M5S 3H4 Canada}
\newcommand\UIUCP{Department of Physics, University of Illinois at Urbana-Champaign, 1110 W. Green Street, Urbana, IL 61801, USA}
\newcommand\UIUCA{Department of Astronomy, University of Illinois at Urbana-Champaign, 1002 W. Green Street, Urbana, IL 61801, USA}
\newcommand\NRAO{National Radio Astronomy Observatory, Charlottesville, NC 22903, USA}
\newcommand\Michigan{Department of Physics, University of Michigan, 450 Church Street, Ann  Arbor, MI 48109, USA}
\newcommand\TorontoP{Department of Physics, University of Toronto, 60 St.\@ George Street, Toronto, ON M5S 3H4 Canada}
\newcommand\Hopkins{Department of Physics and Astronomy, Johns Hopkins University, 3701 San Martin Drive, Baltimore, MD 21218 USA}
\newcommand\Goddard{NASA Goddard Space Flight Center, Code 665, Greenbelt, MD 20771, USA}
\newcommand\APC{APC, Univ. Paris Diderot, CNRS/IN2P3, CEA/Irfu, Obs de Paris, Sorbonne Paris Cit\'e, France}
\newcommand\PennState{Department of Astronomy and Astrophysics, Pennsylvania State University, 520 Davey Lab, University Park, PA 16802, USA}
\newcommand\NIST{National Institute of Standards and Technology, 325 Broadway Mailcode 817.03, Boulder, CO 80305, USA}
\newcommand\Stanford{Department of Physics, Stanford University, 382 Via Pueblo Mall, Stanford, CA 94305, USA}
\newcommand\SLAC{ SLAC National Accelerator Laboratory, 2575 Sand Hill Road, Menlo Park, CA 94025, USA}
\newcommand\PrincetonEngineering{Department of Mechanical and Aerospace Engineering, Princeton University, Engineering Quadrangle, Princeton, NJ 08544, USA}
\newcommand\Fermilab{Fermi National Accelerator Laboratory, P.O. Box 500, Batavia, IL 60510-5011, USA}
\newcommand\KICPChicago{Kavli Institute for Cosmological Physics, University of Chicago, 5640 S Ellis Avenue, Chicago, IL 60637 USA}
\newcommand\Orsay{Institut d'Astrophysique Spatiale, Orsay, France}
\newcommand\MPI{Max-Planck-Institute for Astronomy, Konigstuhl 17, 69117, Heidelberg, Germany}
\newcommand\LAIM{Laboratoire AIM, Paris-Saclay, CEA/IRFU/SAp - CNRS - Universit\'e Paris Diderot, 91191, Gif-sur-Yvette Cedex, France}
\newcommand\WUSTL{Department of Physics, Washington University in St. Louis, 1 Brookings Drive, St.\@ Louis, MO 63130, USA}
\newcommand\MCSS{McDonnell Center for the Space Sciences, Washington University in St. Louis, 1 Brookings Drive, St.\@ Louis, MO 63130, USA}
\newcommand\Austin{Department of Physics, University of Texas, 2515 Speedway, C1600, Austin, TX 78712, USA}
\newcommand\McGill{Department of Physics, McGill University, 3600 Rue University, Montreal, QC, H3A 2T8, Canada}
\newcommand\StewardObs{Steward Observatory, 933 North Cherry Avenue, Tucson, AZ, 85721, USA}
\newcommand\Shahid{Department of Physics, Shahid Beheshti University, 1983969411, Tehran Iran}

\author{ J.~S.-Y.~Leung }
\affiliation{\Toronto}
\affiliation{\TorontoDunlap}

\author{ J.~Hartley }
\affiliation{\TorontoP}

\author{ J.~M.~Nagy }
\affiliation{\WUSTL}
\affiliation{\MCSS}

\author{ C.~B.~Netterfield }
\affiliation{\Toronto}
\affiliation{\TorontoP}

\author{ J.~A.~Shariff }
\affiliation{\CITA}

\author{ P.~A.~R.~Ade }
\affiliation{\Cardiff}

\author{ M.~Amiri }
\affiliation{\UBC}

\author{ S.~J.~Benton }
\affiliation{\Princeton}

\author{ A.~S.~Bergman }
\affiliation{\Princeton}

\author{ R.~Bihary }
\affiliation{\CWRU}

\author{ J.~J.~Bock }
\affiliation{\Caltech}
\affiliation{\JPL}

\author{ J.~R.~Bond }
\affiliation{\CITA}

\author{ J.~A.~Bonetti }
\affiliation{\JPL}

\author{ S.~A.~Bryan }
\affiliation{\ASU}

\author{ H.~C.~Chiang }
\affiliation{\McGill}
\affiliation{\UKZN}

\author{ C.~R.~Contaldi }
\affiliation{\Imperial}

\author{ O.~Dor{\'e} }
\affiliation{\Caltech}
\affiliation{\JPL}

\author{ A.~J.~Duivenvoorden }
\affiliation{\Princeton}

\author{ H.~K.~Eriksen }
\affiliation{\Oslo}

\author{ M.~Farhang }
\affiliation{\Shahid}
\affiliation{\CITA}
\affiliation{\Toronto}

\author{ J.~P.~Filippini }
\affiliation{\UIUCP}
\affiliation{\UIUCA}

\author{ A.~A.~Fraisse }
\affiliation{\Princeton}

\author{ K.~Freese }
\affiliation{\Austin}
\affiliation{\Stockholm}

\author{ M.~Galloway }
\affiliation{\Oslo}

\author{ A.~E.~Gambrel }
\affiliation{\KICPChicago}

\author{ N.~N.~Gandilo }
\affiliation{\StewardObs}

\author{ K.~Ganga }
\affiliation{\APC}

\author{ R.~Gualtieri}
\affiliation{\ANL}

\author{ J.~E.~Gudmundsson }
\affiliation{\Stockholm}

\author{ M.~Halpern }
\affiliation{\UBC}

\author{ M.~Hasselfield }
\affiliation{\PennState}

\author{ G.~Hilton }
\affiliation{\NIST}

\author{ W.~Holmes }
\affiliation{\JPL}

\author{ V.~V.~Hristov }
\affiliation{\Caltech}

\author{ Z.~Huang }
\affiliation{\CITA}

\author{ K.~D.~Irwin }
\affiliation{\Stanford}
\affiliation{\SLAC}

\author{ W.~C.~Jones }
\affiliation{\Princeton}

\author{ A.~Karakci }
\affiliation{\Oslo}

\author{ C.~L.~Kuo }
\affiliation{\Stanford}

\author{ Z.~D.~Kermish }
\affiliation{\Princeton}

\author{ S.~Li }
\affiliation{\Princeton}
\affiliation{\PrincetonEngineering}

\author{D.~S.~Y.~Mak}
\affiliation{\Imperial}

\author{ P.~V.~Mason }
\affiliation{\Caltech}

\author{ K.~Megerian }
\affiliation{\JPL}

\author{ L.~Moncelsi }
\affiliation{\Caltech}

\author{ T.~A.~Morford }
\affiliation{\Caltech}

\author{ M.~Nolta }
\affiliation{\CITA}

\author{ R. O\rq Brient}
\affiliation{\JPL}

\author{ B.~Osherson }
\affiliation{\UIUCP}

\author{ I.~L.~Padilla }
\affiliation{\Toronto}
\affiliation{\Hopkins}

\author{ B.~Racine }
\affiliation{\Oslo}

\author{ A.~S.~Rahlin }
\affiliation{\Fermilab}
\affiliation{\KICPChicago}

\author{ C.~Reintsema }
\affiliation{\NIST}

\author{ J.~E.~Ruhl }
\affiliation{\CWRU}

\author{ M.~C.~Runyan }
\affiliation{\Caltech}

\author{ T.~M.~Ruud }
\affiliation{\Oslo}

\author{ E.~C.~Shaw }
\affiliation{\UIUCP}

\author{ C.~Shiu }
\affiliation{\Princeton}

\author{ J.~D.~Soler }
\affiliation{\MPI}

\author{ X.~Song }
\affiliation{\Princeton}

\author{ A.~Trangsrud }
\affiliation{\Caltech}
\affiliation{\JPL}

\author{ C.~Tucker }
\affiliation{\Cardiff}

\author{ R.~S.~Tucker }
\affiliation{\Caltech}

\author{ A.~D.~Turner }
\affiliation{\JPL}

\author{ J.~F.~van~der~List }
\affiliation{\Princeton}

\author{ A.~C.~Weber }
\affiliation{\JPL}

\author{ I.~K.~Wehus }
\affiliation{\Oslo}

\author{ S.~Wen }
\affiliation{\CWRU}

\author{ D.~V.~Wiebe }
\affiliation{\UBC}

\author{ E.~Y.~Young }
\affiliation{\Stanford}
\affiliation{\SLAC}

\correspondingauthor{Jason S.-Y. Leung}
\email{leung@astro.utoronto.ca}

\begin{abstract}
Modern cosmic microwave background (CMB) analysis pipelines regularly employ complex time-domain filters, beam models, masking, and other techniques during the production of sky maps and their corresponding angular power spectra.
However, these processes can generate couplings between multipoles from the same spectrum and from different spectra, in addition to the typical power attenuation.
Within the context of pseudo-$C_\ell$ based, \texttt{MASTER}-style analyses, the net effect of the time-domain filtering is commonly approximated by a multiplicative transfer function, $F_{\ell}$, that can fail to capture mode mixing and is dependent on the spectrum of the signal.
To address these shortcomings, we have developed a simulation-based spectral correction approach that constructs a two-dimensional transfer matrix, $J_{\ell\ell'}$, which contains information about mode mixing in addition to mode attenuation.
We demonstrate the application of this approach on data from the first flight of the \Spider balloon-borne CMB experiment.
\end{abstract}

\section{Introduction}
\label{sec:intro}

Producing well-characterized maps of the cosmic microwave
background (CMB) from time-ordered data requires accurately accounting for the impact of instrumental effects and any signal processing on the underlying astrophysical signal \citep[\emph{e.g.},][] {jarosik2007three, Planck2015_TOD}.
These processing techniques typically also have a nontrivial impact on the fidelity of the cosmological signal in ways that are spatially anisotropic and inhomogeneous.
These effects must be precisely and accurately characterized in order to avoid biasing the estimation of the CMB angular power spectra, and therefore the inference of cosmological parameters.

The impact of timestream processing and instrument response is commonly approximated in the harmonic-domain with a filter window function unique to the instrument, derived from the analysis of large ensembles of signal simulations.
In its simplest formulation, the processing pipeline is modeled as a power attenuation mechanism in each multipole, \emph{i.e.}, the filter window function is a transfer function -- a simple one-dimensional vector of ratios of output to input power.

This paper addresses the shortcomings of the one-dimensional model and proposes an alternative approach through the construction of a two-dimensional transfer matrix.
This simulation-based spectral correction approach takes into account both the mode mixing and attenuation from instrumental and data processing effects including beams, filtering, and masking.
Section \ref{sec:theory} of this paper describes the theoretical motivation and compares the transfer matrix to the one-dimensional transfer function approach.
A concrete example is presented in Section \ref{sec:application} using data from the first flight of \Spider, a balloon-borne telescope designed to measure CMB polarization on roughly degree angular scales \citep{bmode_paper}.
This section explores different techniques for constructing the transfer matrix to reduce the computational demand including using binned power spectra and performing Fourier-space interpolation.
Section \ref{sec:comparison} presents several comparisons of these techniques, including tests of signal recovery on spectra with different shapes.
While all tested approaches were found to accurately recover a target spectrum identical to that used for the transfer matrix construction, the performance varied when applied to a different target spectrum.
As discussed in Section \ref{sec:conclusion}, this has implications for increasingly sensitive CMB polarization measurements where the cosmological signals are heavily obscured by Galactic foregrounds.
As the foreground power spectra are less well constrained and vary substantially between different sky regions, understanding the signal dependence of potential analysis techniques becomes even more important.

\section{Theoretical Description}
\label{sec:theory}
Maps of the CMB temperature ($T$) and linear polarization ($Q$, $U$) anisotropies over the partial sky can be decomposed into linear combinations of spherical harmonics:
\begin{align}
    T(\textbf{r})W(\textbf{r}) &= \sum\limits_{\ell,m} \tilde{a}^T_{\ell m} Y_{\ell m}(\textbf{r}), \\
    \left[ Q(\textbf{r}) \pm iU(\textbf{r}) \right] W(\textbf{r}) &= \sum\limits_{\ell,m} {}_{\pm 2}\tilde{a}_{\ell m} \; {}_{\pm 2}Y_{\ell m}(\textbf{r}),
\end{align}
where $W(\textbf{r})$ is the window representing the relative weights of the partial-sky mask.
To avoid using spin-weighted spherical harmonic components, the ${}_{\pm 2}\tilde{a}_{\ell m}$ coefficients are frequently expressed as a combination of scalar and fixed parity $E$ and $B$ components
\begin{align}
    {}_{\pm 2}\tilde{a}_{\ell m} = - \left( \tilde{a}^E_{\ell m} \pm i\tilde{a}^B_{\ell m} \right),
\end{align}
where the sign convention follows \cite{zaldarriaga_cmbpol}.
For each $\tilde{a}_{\ell m}$, the pseudo-power spectrum $\tilde{C}_\ell$ is defined as
\begin{align}
    \label{eq:pseudo_cl}
    \tilde{C}_\ell = \frac{1}{2\ell + 1} \sum\limits_{m} \left| \tilde{a}_{\ell m} \right|^2.
\end{align}
Also known as a pseudo-$C_\ell$ (PCL), it is related to the angular power spectrum specified by the theory of primordial perturbations, $C_\ell$, via
\begin{align}
    \label{eq:mode-mode}
    \langle \tilde{C}_\ell^X \rangle = \sum\limits_{\ell'\!\!,\,X'} K_{\ell\ell'}^{XX'} C_{\ell'}^{X'},
\end{align}
where $K_{\ell\ell'}^{XX'}$ is a mode coupling kernel (or ``mixing matrix'') that accounts for the mixing within and between the observables $X \in \{ TT,\, EE,\, BB,\, TE,\, EB,\, TB \}$ due to the partial-sky mask, and the brackets $\langle \cdot \rangle$ denote an ensemble average over infinite spectrum realizations.
Because $K_{\ell\ell'}^{XX'}$ is entirely determined by the chosen pixel weighting and the geometry of the cut sky \citep{master}, in the absence of instrumental effects and noise, PCL estimators can use the (finite) set of measured $\tilde{C}_\ell$ to solve for the underlying power spectrum $C_\ell$.
Examples of PCL estimators include \texttt{MASTER} \citep{master}, \texttt{NaMaster} \citep{NaMaster}, and \texttt{PolSpice} \citep{polspice}.

A major challenge in interpreting CMB data is that the instruments cannot directly probe the true sky signal; the incoming signals are inevitably altered by instrumental systematics and noise.
Therefore, as described in Section \ref{sec:intro}, an experiment's raw datastreams must be processed to remove many different types of spurious signals.
The act of observing and time-domain filtering both distort the signal estimate and present additional sources of mode coupling.
Assuming that the coupling is homogeneous, isotropic, and linear, the impact of the experiment is captured by introducing another coupling matrix, $F_{\ell\ell'}^{XX'}$, such that
\begin{align}
    \label{eq:filter_matrix}
    \langle \tilde{C}_\ell^X \rangle = \sum\limits_{\ell''\!\!,\,X''} \sum\limits_{\ell'\!\!,\,X'} K_{\ell\ell'}^{XX'} F_{\ell'\ell''}^{X'X''} C_{\ell''}^{X''}.
\end{align}
In general, $F_{\ell\ell'}^{XX'}$ is unique to the experiment and cannot be determined analytically.

\subsection{The Filter Transfer Function}
\label{sec:f_ell}
To reduce complexity, $F_{\ell\ell'}^{XX'}$ is often approximated as a diagonal matrix whose entries represent the one-dimensional transfer function $F_\ell^{XX'}$, such that
\begin{align}
    \label{eq:f_ell}
    \langle \tilde{C}_\ell \rangle = \sum\limits_{\ell'} K_{\ell\ell'} F_{\ell'} C_{\ell'}
\end{align}
(here and hereafter, the superscripts $X$ are suppressed for brevity).
Using a set of pseudo-power spectra $\tilde{C}_\ell$ derived from an ensemble of simulations of a known power spectrum $C_\ell$, the transfer function $F_\ell$ can be estimated through an iterative process to avoid inverting $K_{\ell\ell'}$ \citep{master, SPT3G2021}.
Note that $F_\ell$ is frequently decomposed further into a filter component $f_\ell$ and a beam component $b_\ell^2$, such that $F_\ell = f_\ell b_\ell^2$; additional instrumental effects can be inserted in a similar fashion.

This one-dimensional approximation implicitly assumes that each $\ell$-mode remains independent throughout the entire filtering process.
As long as this assumption holds, the mapping from input to output is one-to-one: $F_\ell$ is simply the ratio of output to input power for each $\ell$.
However, in practice, we expect modes to become coupled with one another, where the mapping becomes many-to-one and the contributions from the coupled input modes become impossible to disentangle using $F_\ell$ alone.
The consequence is that changing the input power spectrum $C_\ell$ also changes the output $\tilde{C}_\ell$ in some nontrivial way due to this many-to-one mapping.
In other words, $F_\ell$ is inextricably tied to the input used to compute it.

More concretely, the one-dimensional transfer function formulation conflates mode mixing with the in-mode filter gain.
This introduces a sensitivity to the power spectrum of the simulated sky used to calibrate $F_\ell$; the final spectrum is correct only if the simulated sky is statistically similar to the true sky (\emph{i.e.}, a Gaussian sky realization with the same power spectra).

\begin{figure*}[tbp]
    \centering
    \includegraphics[width=\textwidth]{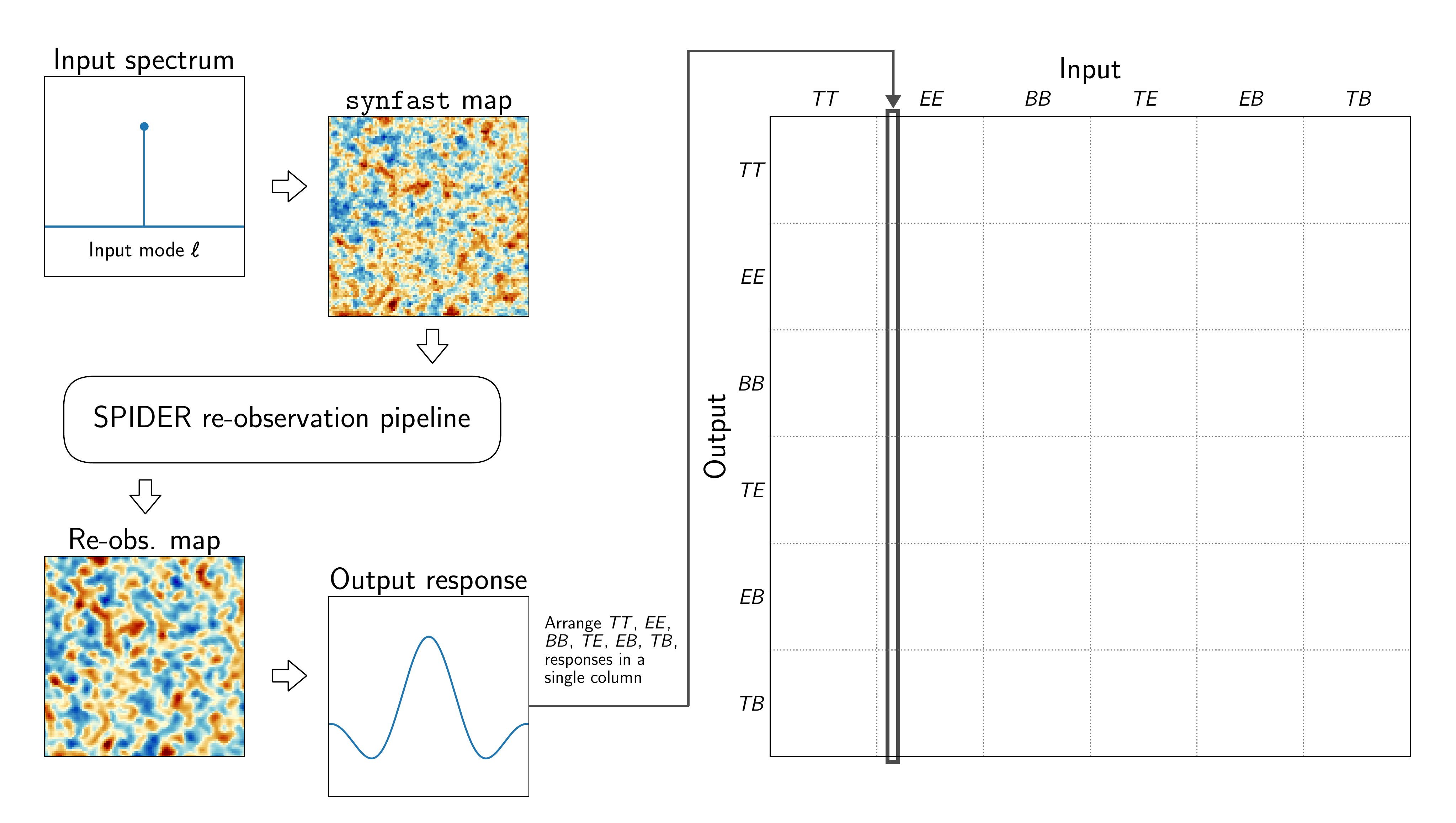}
    \caption{Procedure to create the transfer matrix $J_{\ell\ell'}$. For each input mode, the six output response spectra (gains) are arranged vertically to form the columns of the matrix. Each of the 36 blocks within the matrix spans a range of multipoles with $\ell_\text{min} \le \ell \le \ell_\text{max}$.}
    \label{fig:layout}
\end{figure*}

\subsection{The Multipole--Multipole Transfer Matrix}
To address this shortcoming of the $F_\ell$ formulation, we introduce a two-dimensional linear operator that encodes the (asymmetric) coupling between each $\ell\ell'$ pair:
\begin{align}
    \label{eq:jll_def}
    \langle \tilde{C}_\ell \rangle &= \sum\limits_{\ell'} J_{\ell\ell'} C_{\ell'}.
\end{align}
We refer to this coupling operator $J_{\ell \ell'}$ as the ``transfer matrix'' because it directly relates the input of the true power spectrum to the output of the spectrum estimator.
This relation holds as long as the filtering process is approximately linear.
Treating the entire pipeline as a single operator avoids the diagonal approximation and ensures that all multipole--multipole couplings induced by time-domain and map-domain filtering are properly taken into account, \emph{i.e.}, we are not locked into a specific input spectrum.

Note that these couplings extend to those between the six power spectra; both temperature-to-polarization ($T$-to-$P$) and $E$-to-$B$ leakage are automatically included.
Because standard PCL spectrum estimators provide the $TT$, $EE$, $BB$, $TE$, $EB$, and $TB$ power spectra, any input mode can be readily related to an output mode from any of these six power spectra, resulting in $6^2$ coupling matrices.
We find it convenient to compile these individual matrices into a single $6 \times 6$ block matrix encapsulating every $\ell\ell'$ coupling between the six power spectra.
Following the rules of matrix multiplication (Equation \ref{eq:jll_def}), we arrange these blocks horizontally according to input spectra and vertically according to output spectra (see Figure \ref{fig:layout}).
The ordering of the six spectra does not matter as long as it is consistent between the two axes; likewise, the six $C_\ell$s must be concatenated in the same order.
We choose the above ordering based on convenience.

While the approach presented here has similarities to that used by the BICEP/Keck Collaboration for their CMB polarization analyses, the implementation varies due to key differences in the observing strategies and analysis pipelines.
As described in \cite{BK_observation_matrix}, the simplicity of the BICEP/Keck observing strategy allows for the construction of an observing matrix that renders large simulation ensembles more computationally tractable.
Having determined their filtering operations to be linear, the observing matrix is used to construct the transfer matrix $J_{\ell\ell'}$, which is then used similarly to reconstruct and interpret the on-sky power spectra.
Because the observing matrix formulation is less generally applicable to experiments at other sites with different observing strategies, the remainder of this paper is dedicated to estimating $J_{\ell\ell'}$ through other means.

\section{Application to \Spider Data}
\label{sec:application}
To illustrate the application of the transfer matrix, we present results from the \Spider balloon-borne telescope.
During its first Antarctic long-duration balloon (LDB) flight in January 2015, \Spider mapped \SI{4.8}{\percent} of the sky with polarimeters operating at 95 and \SI{150}{\giga\hertz} to constrain the $B$-mode power spectrum from primordial gravitational waves \citep{bmode_paper}.
An upcoming flight with additional \SI{280}{\giga\hertz} receivers will provide improved characterization of the polarized Galactic dust foregrounds \citep{shaw_spie}.

\Spider's processing pipeline is described more fully in \cite{bmode_paper} and \cite{tod_forward};
here we briefly highlight the most relevant steps and note that they are sufficiently linear to allow usage of Equation \ref{eq:jll_def}.
Once features such as cosmic-ray hits, payload transmitter signals, and thermal transients have been removed from the raw detector timestreams, they are filtered to reduce quasi-stationary noise.
Null test performance was used to identify the weakest filter that sufficiently removed quasi-stationary noise, resulting in a fifth-order polynomial fit to each detector's data as a function of azimuth angle between scan turnarounds.
The impact of scanning, filtering, and flagging are determined by applying the entire processing pipeline to an ensemble of time-domain signal simulations in a procedure known as re-observation.
The re-observed timestreams are produced at the full data sample rate without any downsampling or binning.
Unlike the BICEP/Keck experiment, \Spider's observing strategy does not allow for the creation of an observing matrix, so obtaining the transfer matrix $J_{\ell\ell'}$ requires producing an appropriate set of re-observed CMB maps.

\Spider's measured and simulated timestreams are converted into two-dimensional maps of the sky with a binned mapmaker.
This approach assembles the detector data into spatial pixels based on the telescope pointing and polarization sensitivity as described further in \cite{bmode_paper}.
The computational simplicity of this method is crucial for enabling the generation of large simulation ensembles, including those used in this work.

The main \Spider cosmological results presented in \cite{bmode_paper} use two complementary pipelines for power spectrum estimation.
Here we describe results only from the Noise Simulation Independent (NSI) pipeline, while the other pipeline is presented in \cite{xfaster_forward}.
The NSI pipeline is similar to Xspect \citep{Xspect} and Xpol \citep{Xpol}, as well as to those used by SPT \citep{lueker_SPT}, CLASS \citep{padilla_class}, and \Spider's circular polarization analysis \citep{SpiderVpol}.
It decomposes the 2015 flight data into 14 maps by splitting the timestreams into interleaved 3-minute segments for each of the six receivers.
This timescale was chosen to maximize the number of complete and well-conditioned maps of the sky region, taking into account \Spider's scan rate and observing strategy.
These maps are then co-added by frequency to produce 14 maps in each of the two \Spider bands, 95 and \SI{150}{\giga\hertz}.
The cross-power spectra for all pairs of maps (neglecting the auto-spectra) are produced with \texttt{PolSpice} \citep{polspice}, and the distribution of the cross-spectra allows for an empirical measurement of the uncertainty without reliance on an instrumental noise model.
In \Spider's NSI pipeline, these cross-spectra are used to estimate cosmological parameters by comparing them to theoretical models that depend on the parameters of interest.
It is therefore \Spider's power spectra, rather than the maps, that we correct for the beam and filtering effects as described in the rest of this work.
Since the \texttt{PolSpice} estimator is linear, the assumption of linearity in the definition of the transfer matrix (Equation \ref{eq:jll_def}) is satisfied.
Although the transfer matrix can be used to correct the power spectra for temperature-to-polarization leakage, we perform this correction on the maps based on simulations of \Planck temperature-only maps \citep{planck18_hfi}.
This avoids the sample variance from the simulation ensemble, which is relatively large due to the amplitude of the temperature signals, by using existing measurements of the temperature anisotropies in \Spider's observing region.

As $J_{\ell\ell'}$ is merely the linear operator linking the input and output power spectra (Equation \ref{eq:jll_def}), its computation is conceptually straightforward: pass a (simulated) CMB map with a known spectrum through the re-observation pipeline, then compute the power spectra of the output map.
But to capture the asymmetric mode--mode coupling between each $\ell\ell'$ pair, this process must be performed on each mode individually.
Thus we begin with a set of unit $\delta$-functions -- one at each $\ell$ of interest -- and simulate a set of CMB maps using the \texttt{synfast HEALPix} utility\footnote{\url{https://healpix.sourceforge.io }} \citep{HealPix}.
Upon obtaining the re-observed maps, we organize their power spectra into columns, arranged by $\ell$, to form $J_{\ell\ell'}$.
The whole process is illustrated schematically in Figure \ref{fig:layout}.

To reduce the effects of sample variance from the CMB map realization and partial-sky spectrum estimation procedures, we repeat the above process 100 times and average the 100 resulting matrices to obtain the final $J_{\ell\ell'}$.
However, the computation cost is intensive: with 100 map realizations of a $\delta$-function in each of the three CMB modes ($TT$, $EE$, $BB$) that must be re-observed by each of \Spider's six receivers, we require 1800 simulated maps per $\ell \in [2,250+\Delta\ell_b]$ (a $\Delta\ell_b$ ``buffer'' is necessary to account for couplings of higher multipoles with lower multipoles).
To minimize the computational burden, we use full-mission \Spider maps instead of the 14 NSI pipeline submaps because this was demonstrated to have a negligible effect on \Spider's signal recovery in simulations.
Nevertheless, re-observing the entire set of required maps would take about 1000~core-years on the \texttt{Niagara} supercomputing cluster \citep{Scinet,Niagara}, which is computationally infeasible given \Spider Collaboration resources.
Since the fidelity of the computation needs to be balanced against the time required, two practical options are investigated in the following sections: 1) use input spectra with multipole bins rather than evaluating each multipole individually, or 2) compute the transfer matrix for only a sample of input $\ell$s and interpolate between them.

\begin{figure*}[tbhp]
    \centering
    \includegraphics[width=\textwidth]{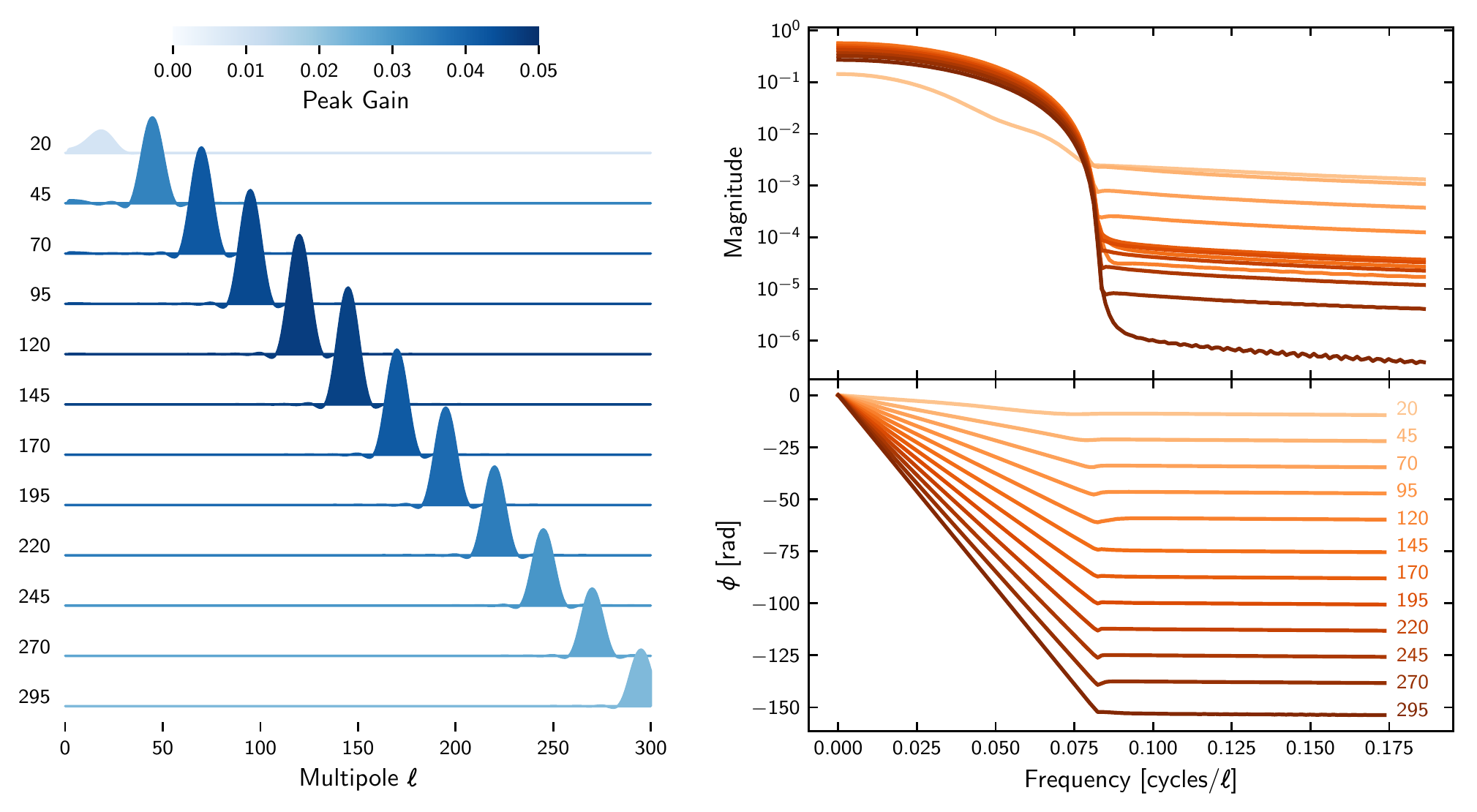}
    \caption{(\emph{left}) Impulse responses of unit $\delta$-function inputs at $\ell = {20, 45, \dots, 295}$ for the \SI{150}{\giga\hertz} $EE \rightarrow EE$ coupling; the $TT$ and $BB$ counterparts are similar. (\emph{right}) The corresponding magnitude and (unwrapped) phase responses. The ``frequency'' axis has units cycles/$\ell$ and does not have a simple physical interpretation. These responses are interpolated in the Fourier domain to approximate the transfer matrix (Figure \ref{fig:block}).}
    \label{fig:fourier}
\end{figure*}

\begin{figure*}[tbhp]
    \centering
    \includegraphics[width=\textwidth]{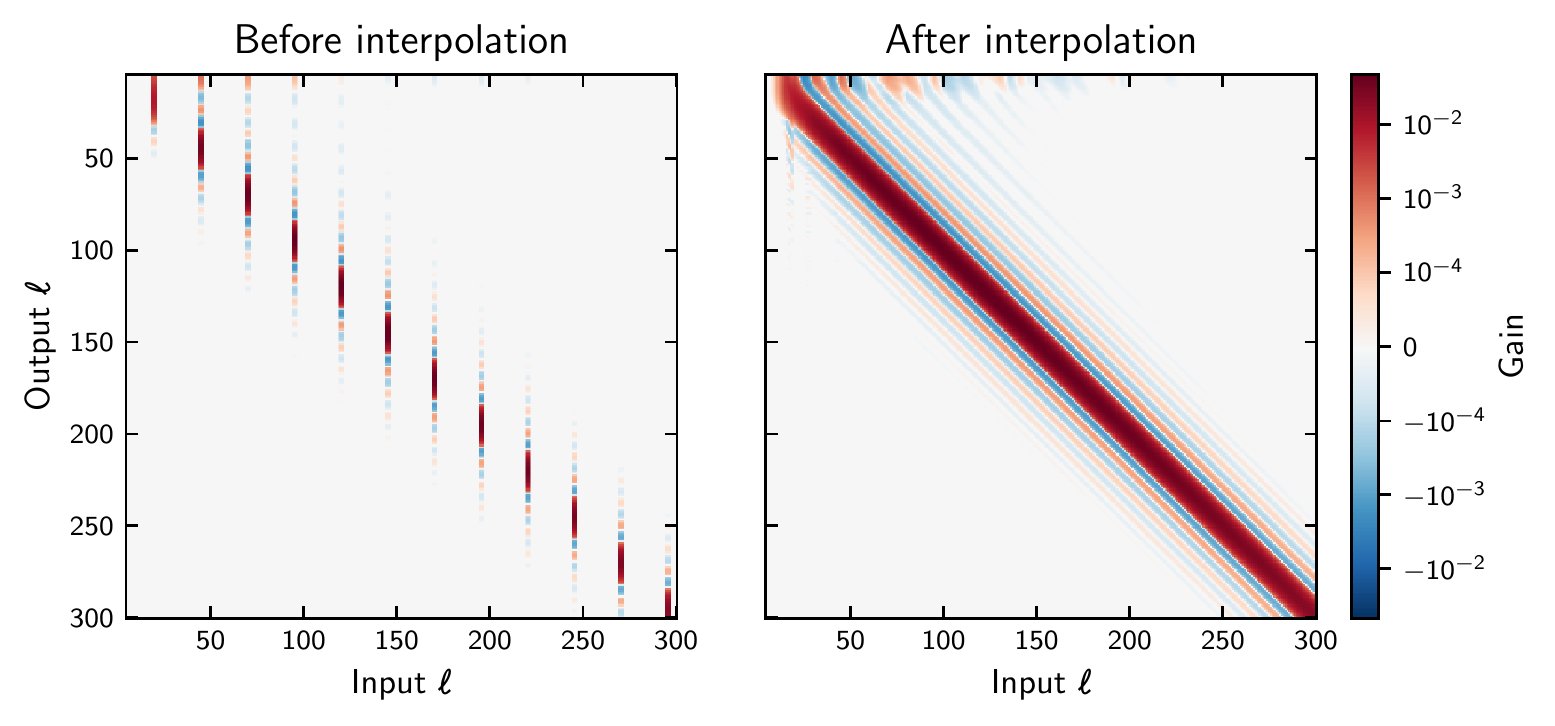}
    \caption{The impulse responses from the left panel of Figure \ref{fig:fourier}, arranged into columns (\emph{left}), are interpolated in Fourier space to form an approximated $J_{\ell\ell'}^\text{interp}$ (\emph{right}). Shown is the \SI{150}{\giga\hertz} $EE \rightarrow EE$ coupling; the \textit{TT} and \textit{BB} counterparts are similar.}
    \label{fig:block}
\end{figure*}

\subsection{Bin--Bin Transfer Matrix}
CMB angular power spectra are typically binned into multipole ranges of width $\Delta\ell$ in order to increase the signal-to-noise ratio and mitigate the correlative effects of adjacent multipole leakage.
As a result, we consider the case where the input power spectra are also binned into the same multipole ranges, thereby reducing the computational demands of the calculation.
By analogy with Equation~\ref{eq:jll_def}, we define a bin--bin transfer matrix $J_{bb'}$ that describes the average power in output bin $b$ given the average power in each of the input bins $b'$:
\begin{align}
    \label{eq:jbb_def}
    \langle \tilde{C}_b \rangle = \sum\limits_{b'} J_{bb'} \sum\limits_{\ell} P_{b'\ell} C_{\ell} = \sum\limits_{b'} J_{bb'} C_{b'},
\end{align}
where $P_{b\ell}$ is a binning operator.

At first glance, the $\delta$-function inputs to $J_{\ell\ell'}$ might be replaced by unit-boxcars for $J_{bb'}$, but, as it turns out, that may not be the best choice.
Unlike the case for the full $J_{\ell\ell'}$, the binning introduces a dependence on the shape of the input spectrum used to estimate the transfer matrix (see the \hyperref[sec:appendix]{Appendix}).
Consequently, the choice of input spectrum should be as close as possible to the anticipated output spectrum, as is further discussed in Section \ref{sec:comparison}.
Because our application is toward degree-scale CMB $B$-modes, we use a $\Lambda$ cold dark matter $\text{(CDM)} + (r = 0.03)$ spectrum, provided by \texttt{CAMB} \citep{camb}, as our base model.
Windowing this model spectrum at each bin of interest -- \emph{i.e.}, multiplying it by unit-boxcars centred on each bin -- provides the set of input spectra needed to construct the bin--bin transfer matrix $J_{bb'}$.

Because the analysis of \Spider data uses only 12 bins in the range $2 \le \ell \le 300$, the computational demand is decreased by a factor of $\sim$30; months of cluster time is reduced to days.
The price is a dependence on an input model, but unlike $F_\ell$, the ability of $J_{bb'}$ to capture couplings between multipole bins provides a more general framework.

\subsection{Fourier-space Interpolation}
\label{sec:interp}
An alternative method for reducing computation time is to compute the transfer matrix $J_{\ell\ell'}$ only at select multipoles (\emph{i.e.}, columns) and interpolate in between.
This requires the matrix to be smooth in the $\ell$-range of interest, but, as we shall see, this assumption is often well justified in practice.

Because $J_{\ell\ell'}$ is a linear operator (Equation \ref{eq:jll_def}), the output from a $\delta$-function input can be interpreted as the impulse response of a linear filter.
As shown in the left panel of Figure \ref{fig:fourier}, \Spider's transfer components ($TT \rightarrow TT$, $EE \rightarrow EE$, etc.) have impulse responses resembling displaced sinc functions.
This allows us to recast the problem from the $\ell$ domain to its dual frequency-like domain (hereafter simply the ``frequency'' domain), where the interpolation turns out to be considerably easier.
For our application, we use the sign and normalization conventions established by NumPy's \texttt{fft} module;\footnote{\url{https://numpy.org/doc/stable/reference/routines.fft.html}} for a given input $\ell'$, the discrete Fourier transform of its impulse response $x_\ell$ -- the frequency response -- is defined by:
\begin{align}
    X_f = \sum\limits_{\ell=0}^{N-1} x_\ell \exp\left( -i \frac{2\pi f}{N} \ell \right).
\end{align}
We use $N=801$ points ($0 \le \ell \le 800$) in our computations.
The magnitudes $|X_f|$ and phases $\phi = \angle X_f$ of the resulting transforms are shown in the right panel of Figure \ref{fig:fourier}.

In general, the properties of the responses are unique to each experiment and therefore often determined empirically.
\Spider's response corresponds to a linear-phase low-pass filter: the magnitude rolls off until it hits the noise floor of the filter at $f = f_c$, while the phase $\phi$ is nearly linear up to that same frequency.
The slope of the linear-phase component is proportional to the shift (in $\ell$) of the impulse, namely, $m = \phi/f = -2\pi\ell$ when $f$ is expressed in normalized frequency units (cycles per multipole).
The magnitude of the noise floor encodes the level of background noise at the input multipole $\ell'$.

This simple structure in the frequency domain permits a straightforward interpolation.
The magnitudes, in particular, can be interpolated directly at all $f$, as can the unwrapped phases at $f < f_c$.
For simplicity, we interpolate piecewise-linearly.
In the noise region $f > f_c$, the \textit{wrapped} phases exhibit a curious behaviour: they settle onto one of two possible values separated by $\pi$.
Thus, rather than interpolating, we implement a nearest-neighbour scheme for the phases at $f > f_c$.

\begin{figure}[tbp]
    \centering
    \includegraphics[width=0.47\textwidth]{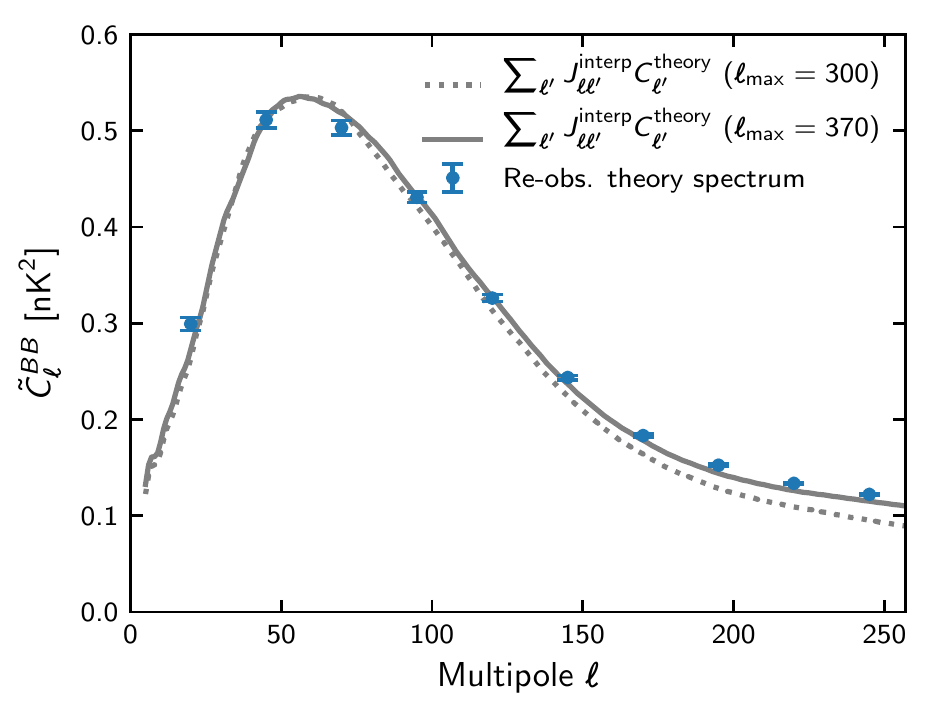}
    \caption{Although \texttt{anafast} does not automatically correct for $E$-to-$B$ leakage, a transfer matrix built using this utility can still capture such mode mixing induced by \Spider's filter and mask. When the matrix is applied to a $\Lambda\text{CDM} + (r = 0.03)$ theory spectrum, the result can predict the output from re-observing a map with the same theory spectrum (\emph{blue points}; error bars indicate the error on the mean of the 500 simulations). However, replicating the full effects of the re-observation procedure in the higher bins requires computing a larger matrix than anticipated due to the wide tails of the $EE \rightarrow BB$ response. Namely, a matrix with $\ell_\text{max} = 300$ (\emph{dotted line}) underestimates the power in the higher bins; this is alleviated somewhat by increasing to $\ell_\text{max} = 370$ (\emph{solid line}), but that is still not quite enough. As a result, we proceed with a matrix constructed using the \texttt{PolSpice} estimator, as its built-in $E$-to-$B$ leakage correction algorithm avoids the need to generate larger matrices, thereby reducing the computational demand.}
    \label{fig:e-to-b}
\end{figure}

\begin{figure}[t]
    \centering
    \includegraphics[width=0.47\textwidth]{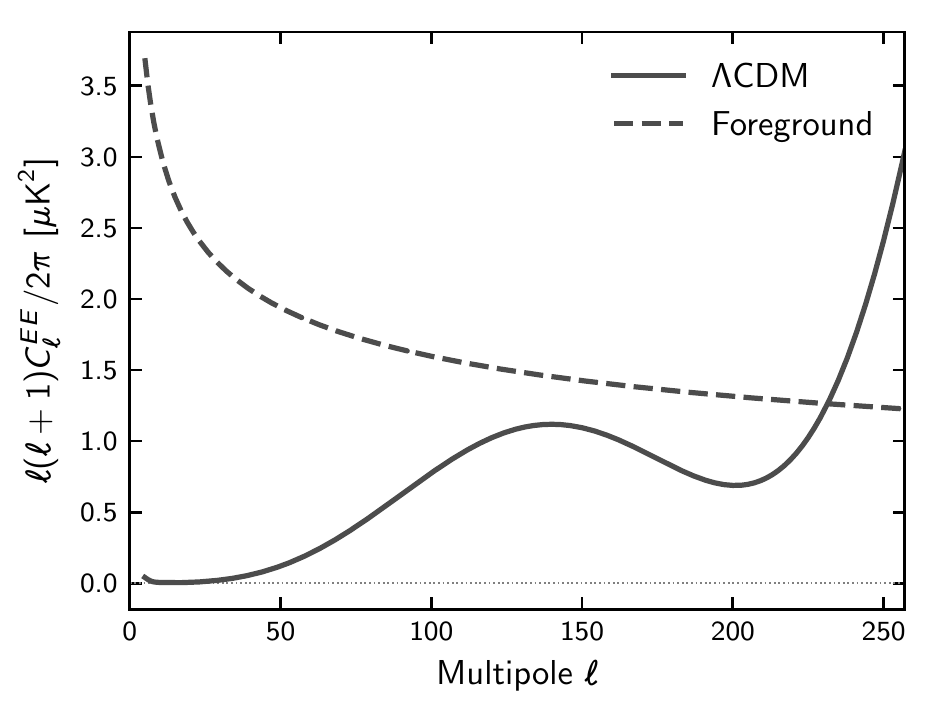}
    \caption{Two fiducial input theory spectra with very different shapes are used to evaluate the performance of different methods of computing the transfer function: a theoretical $\Lambda$CDM CMB spectrum and a power-law Galactic foreground model.  Here only the CMB $EE$ spectrum is shown, while the same foreground model is used for temperature and polarization.}
    \label{fig:theory_input}
\end{figure}

\begin{figure*}[t]
    \centering
    \includegraphics[width=0.8\textwidth]{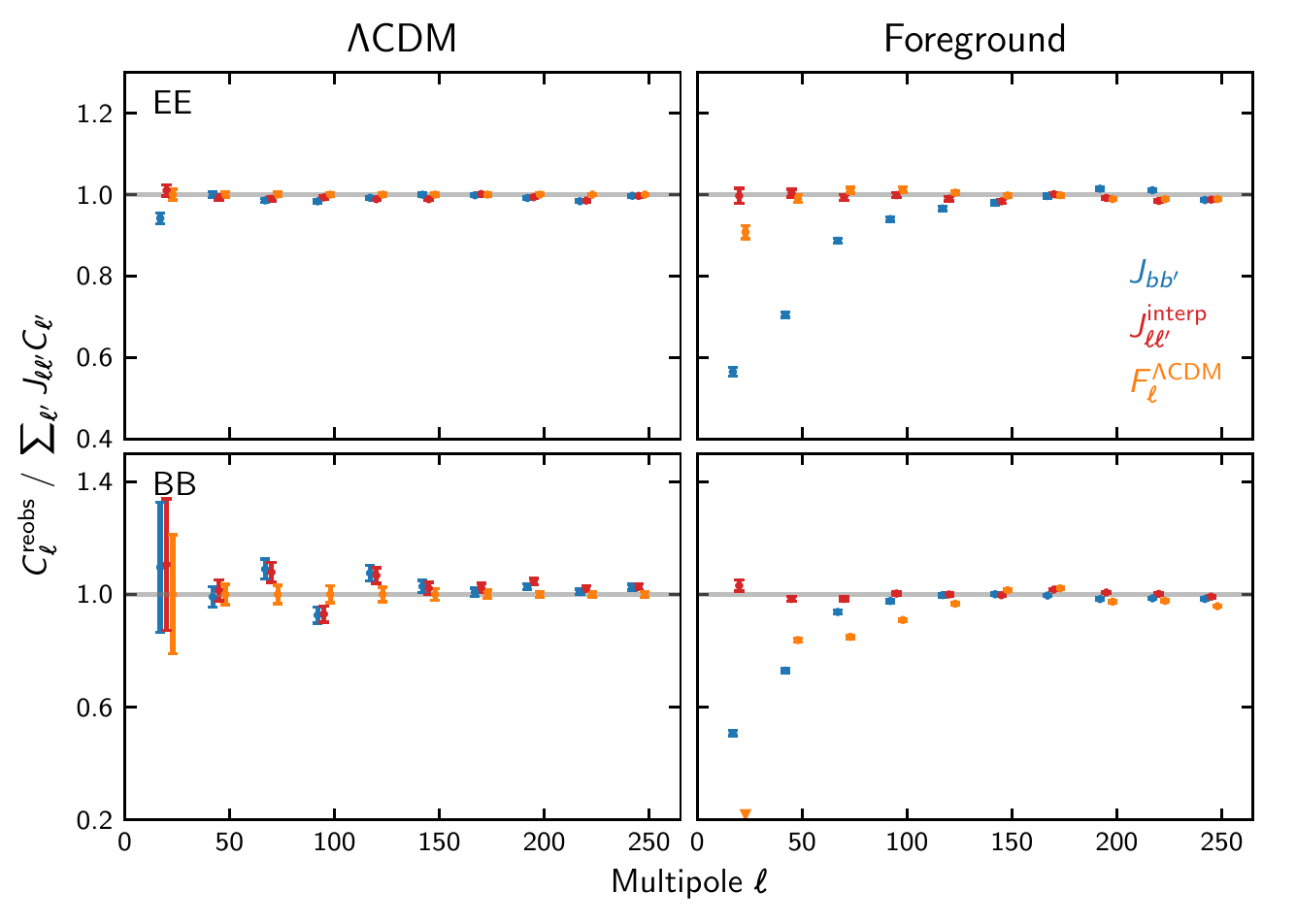}
    \caption{Average ratios of re-observed 150~GHz spectra to those obtained from applying the transfer function and transfer matrices to fiducial $\Lambda\text{CDM} + (r = 0.03)$ and foreground input spectra (500 and 300 realizations, respectively). A ratio of 1 represents the ideal performance, signifying a perfect replication of \Spider's filtering process. As the transfer function $F_\ell$ was derived using the re-observed $\Lambda$CDM result, applying it to that same spectrum again returns 1 by definition (\emph{left panels}). Error bars indicate the error on the mean of the simulation ensembles.}
    \label{fig:xferratios}
\end{figure*}

\begin{figure*}[tbp]
    \centering
    \includegraphics[width=\textwidth]{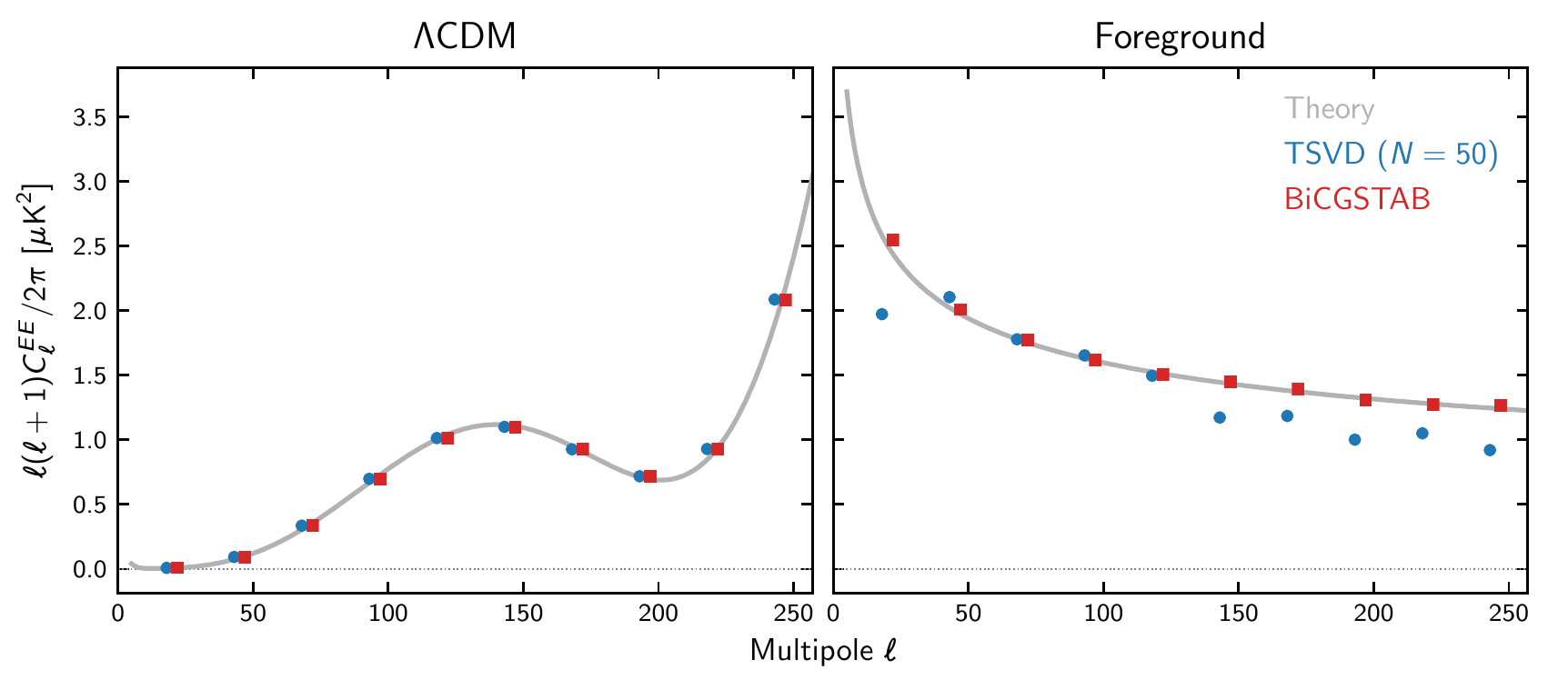}
    \caption{Solutions to the linear system $\sum_{\ell'} J_{\ell\ell'}^\text{interp} x_{\ell'} = \sum_{\ell'} J_{\ell\ell'}^\text{interp} C_\ell^\text{theory}$ for a $\Lambda$CDM $EE$ spectrum (\emph{left}) and a dust foreground spectrum (\emph{right}) computed by the truncated singular value decomposition (TSVD) method with 50 singular values and the biconjugate gradient stabilized (BiCGSTAB) method. Near-optimal results can be obtained when using the TSVD matrix as a preconditioner and providing an initial guess close to $C_\ell^\text{theory}$ (details provided in the text). Though not shown here, $TT$ and $BB$ spectra return similar results.}
    \label{fig:solveinv}
\end{figure*}

One might be tempted to filter out the noise at $f > f_c$ completely.
This requires care: any operation in the frequency domain carries consequences into the $\ell$ domain.
For example, a simple low-pass filter with a rigid cutoff is equivalent to convolution with a sinc function in $\ell$-space; this operation will pollute the impulse response with heavy ringing, and should be avoided.
Because good filter design merits a study of its own, we do not carry out any filtering of this sort.

To complete the process, the interpolated values are inverse Fourier transformed back into $\ell$-space to form our interpolated $J_{\ell\ell'}^\text{interp}$ (Figure \ref{fig:block}).
The choice of nodes for the interpolation, \emph{i.e.}, which columns of $J_{\ell\ell'}$ to compute rather than interpolate, reduces to a balance between fidelity and computation time.
Naturally we will want to concentrate our efforts to regions where $J_{\ell\ell'}$ changes rapidly in order to obtain a good interpolation.
For us, that is the low-$\ell$ regime.
We thus compute the columns at $\ell = \{5, 10, 15\}$ in addition to \Spider's bandcenters.

We note that this particular choice of interpolation method is motivated by \Spider's sinc-function-like impulse responses.
However, any sufficiently smooth $J_{\ell\ell'}$ can be interpolated with similar techniques.

\subsection{Choice of Spectrum Estimator}
\label{sec:choice}
The matrix-making process encodes the spectrum estimator into the matrix -- that is, $J_{\ell\ell'}$ depends on the choice of estimator by design (Equation \ref{eq:jll_def}).
Nonetheless, this choice can be trivially changed because the spectrum estimation and re-observation steps are completely independent of each other.

The \texttt{anafast} utility (provided in the \texttt{healpy} package) is a simple spectrum estimator that directly computes the power spectra from a set of input maps using Equation \ref{eq:pseudo_cl}.
In general, the interpolated matrix constructed using \texttt{anafast} provides very good agreement with the expected results.
However, we find that the ``buffer'' needs to be quite large ($\Delta\ell_b > 120$) in order to properly account for $E$-to-$B$ leakage (Figure \ref{fig:e-to-b}).
To avoid the additional computational demand, in the following work, we proceed using the \texttt{PolSpice} estimator, which includes as part of its algorithm a correction for $E$-to-$B$ leakage built in.
This reduces the ``buffer'' to a much more manageable $\Delta\ell_b = 50$.

We use the same \texttt{PolSpice} parameters as the \Spider $B$-mode analysis for this work, including a cosine apodization with $\sigma = \SI{10}{\degree}$, $\theta_\text{max} = \SI{50}{\degree}$, and \texttt{symmetric\_cl} \citep{bmode_paper}.

\subsection{Cross-spectra: $TE$, $EB$, and $TB$}
An outstanding problem in our discussion so far is the treatment of the cross-spectra $TE$, $EB$, and $TB$.
While the inter-spectral couplings from auto- to cross-spectra ($TT \rightarrow TE$, etc.) are implicitly accounted for by the spectrum estimator (which returns all six power spectra), the method described here cannot be used to compute the matrix components with a cross-spectrum as input.
To wit, it is impossible to generate maps of Gaussian-distributed $a_{\ell m}$s with a nonzero correlation (the input mode) in $TE$, $EB$, or $TB$ while the $TT$, $EE$, and $BB$ spectra are imposed to be zero.
Consequently, we approximate the diagonal cross-spectra transfer components ($TE \rightarrow TE$, etc.) as the geometric mean of the related auto-spectra components, $J_{\ell\ell'}^{TE,TE} = (|J_{\ell\ell'}^{TT,TT}| |J_{\ell\ell'}^{EE,EE}|)^{1/2}$, etc., and treat the off-diagonal components ($TE \rightarrow EB$, etc.) as negligible.
Although this is not ideal, it was found not to bias simulation results for our application.
An alternative method could be to construct the transfer matrix using the harmonic modes, rather than the power spectra, of the maps.
As an example, see \cite{ACT2020}, who postulate a transfer matrix built from (two-dimensional) Fourier transforms of $T$, $E$, and $B$ maps.

\section{Comparison of Methods}
\label{sec:comparison}
We quantify the effectiveness of the two methods described above (binning and interpolating) by two metrics: their ability to replicate the end-to-end \Spider pipeline, and to recover the original signal.
In the context of Equation \ref{eq:jll_def}, this corresponds to producing the correct $\tilde{C}_\ell$ when given ${C}_\ell$, and vice versa.

\subsection{Case 1: Solve for $\tilde{C}_\ell$, Given $C_\ell$}
We demonstrate the first of these by applying our computed transfer function/matrices directly to a theory spectrum.
This result is then compared to the output spectrum generated by passing a set of map realizations of the same input theory spectrum through the \Spider re-observation procedure.
To ensure broad applicability, we use two qualitatively different theory spectra as input: a $\Lambda\text{CDM} + (r = 0.03)$ spectrum (from \texttt{CAMB}; 500 realizations) and a power-law dust foreground spectrum (300 realizations).
These two spectra are chosen for their contrasting behaviour (Figure \ref{fig:theory_input}).

As discussed in Section \ref{sec:f_ell}, the input-dependence of $F_\ell$ makes it suitable only when the target spectrum does not deviate far from the input spectrum that was used to construct $F_\ell$.
This is demonstrated in Figure \ref{fig:xferratios}.
Having derived our $F_\ell$ using a fiducial $\Lambda\text{CDM} + (r = 0.03)$ spectrum as input, applying it to the same spectrum again outputs a trivial result (left panels).
But when the target is the foreground spectrum (right panels), large deviations occur wherever mode--mode coupling dominates over in-mode gain -- \emph{i.e.,} wherever the transfer matrix is least diagonal-like.
In our case, this occurs at low $\ell$s (Figure \ref{fig:block}).
The simple power attenuation model of $F_\ell$ is insufficient in capturing \Spider's filtering scheme in this regime.

The same kind of input-dependence is seen in the binned transfer matrix $J_{bb'}$ (see the \hyperref[sec:appendix]{Appendix}).
Like $F_\ell$, the $J_{bb'}$ constructed from a fiducial $\Lambda\text{CDM} + (r = 0.03)$ spectrum is highly effective when the target spectrum is also a $\Lambda\text{CDM} + (r = 0.03)$ spectrum, and diverges when the target is the foreground dust spectrum (Figure \ref{fig:xferratios}).
But this time, all foreground bins up to $\ell = 100$ are noticeably afflicted.
This is due to the loss of granularity from binning the spectrum and the matrix, as we had used a bin width of $\Delta\ell = 25$ to match \Spider's $B$-mode analysis.
In the limit where the bin width is 1 -- a single multipole -- we recover the full, theoretically ideal transfer matrix $J_{\ell\ell'}$.

The interpolated transfer matrix $J_{\ell\ell'}^\text{interp}$ thus functions as a compromise allowing us to have a bin width of 1 without being computationally intractable.
Despite the simple piecewise-linear interpolation approximation used, it reliably recovers both target spectra.

\subsection{Case 2: Solve for $C_\ell$, Given $\tilde{C}_\ell$}
To recover $C_\ell$ from $\tilde{C}_\ell$, the linear system defined by Equation \ref{eq:jll_def} must be solved numerically; directly inverting a linear system is not recommended except in special cases.
One such case is the transfer function $F_\ell$ -- computing its inverse is trivial since it is a diagonal matrix (a one-dimensional function).
Likewise, \Spider's binned transfer matrix $J_{bb'}$ is also easily invertible because it is diagonally dominant and measures only $72 \times 72$ (12 bins in each of the six correlation spectra); its condition number is very close to 1.
The main challenge, therefore, is solving the system for the interpolated transfer matrix $J_{\ell\ell'}^\text{interp}$, whose condition number is $\mathcal{O}(10^{17})$.

The large condition number of $J_{\ell\ell'}^\text{interp}$ -- and also of $J_{\ell\ell'}$ in general -- means that it is highly susceptible to numerical noise and cannot be solved using elementary linear algebraic techniques.
Numerous algorithms of varying complexities exist to solve such ill-conditioned linear systems; the study of these techniques is beyond the scope of this paper.
As a demonstration, below we focus on the truncated singular value decomposition (TSVD) method and the biconjugate gradient stabilized (BiCGSTAB) method \citep{bicgstab}.
To test the efficacy of these two techniques, we compute $\tilde{C}_\ell = \sum_{\ell'} J_{\ell\ell'}^\text{interp} C_\ell^\text{theory}$, solve the system $\sum_{\ell'} J_{\ell\ell'}^\text{interp} x_{\ell'} = \tilde{C}_\ell$, and compare the solution $x_\ell$ with the original $C_\ell^\text{theory}$.
The results are shown in Figure \ref{fig:solveinv}.
We use the same test spectra as the previous section (Figure \ref{fig:theory_input}) as $C_\ell^\text{theory}$.

TSVD approximates the inverse matrix by rank-reducing the factorized matrices such that only the $N$ largest singular values remain.
Because singular values represent magnitudes along orthogonal axes, this truncation removes the low-amplitude components most susceptible to numerical instabilities.
For our $J_{\ell\ell'}^\text{interp}$, we truncate the singular values at the quasi-arbitrary cutoff of $10^{-3}$, which leaves the top 50 singular values; below this cutoff, the singular vectors are inundated with numerical noise.
As shown in Figure \ref{fig:solveinv}, TSVD with $N = 50$ does well for a $\Lambda$CDM spectrum but remains noisy for a dust power law.

On the other hand, BiCGSTAB is an iterative algorithm.
While we find that its standard configuration can recover the $C_\ell^\text{theory}$ to an acceptable tolerance, its convergence can be aided by a suitable preconditioner matrix and initial guess.
For convenience, we use the truncated matrix from our TSVD tests above as the preconditioner, and, to replicate real conditions, we use a noisy version of $C_\ell^\text{theory}$ ($\pm$\SI{5}{\percent} noise) as the initial guess.
This produces near-optimal results (Figure \ref{fig:solveinv}) in fewer than 10 iterations.

\section{Conclusion and discussion}
\label{sec:conclusion}
The transfer function $F_\ell$ is an \textit{Ansatz} introduced by \citet{master} to represent the effects of datastream filtering in the process of CMB mapmaking.
It was noted at the time that $F_\ell$ cannot fully capture such filtering, on account of inconsistencies with the statistics governing the power spectrum $C_\ell$ (in particular, time-domain filtering violates the assumption of isotropy), but the authors showed that such inconsistencies had a negligible impact on their results.
With filtering and analysis techniques having grown far more complex as CMB experiments have become more sensitive in the two decades since, and considering the crucial role $F_\ell$ fills in recovering cosmological signals, this \textit{Ansatz} merits a more thorough re-examination.
This is particularly important for measurements of CMB polarization in the presence of large foreground components, where there is significant uncertainty in the shape of the power spectrum for a given sky region, and where foreground residuals may be a concern in component-separated maps.

In this study, we present a different approach, the transfer matrix $J_{\ell\ell'}$, that accounts for a key element missing in $F_\ell$: the multipole--multipole coupling induced by the instrument response and timestream filtering.
These couplings are not restricted to those within the same spectrum; also included are inter-spectral coupling components such as $T$-to-$P$ and $E$-to-$B$ leakage.
Because the full matrix is computationally infeasible for most modern experiments, we additionally explore two approximations: a binned transfer matrix $J_{bb'}$ and an interpolated transfer matrix $J_{\ell\ell'}^\text{interp}$.
Along with $F_\ell$, we compare their performance by applying them to the sky-to-spectrum pipeline from \Spider's first flight.

When the target is a $\Lambda$CDM-like spectrum, we find that all three approaches are effective at replicating the sky-to-spectrum pipeline.
However, when the target is switched to a power-law dust foreground model, $F_\ell$ and $J_{bb'}$ -- both being dependent on the fiducial model used to construct them -- falter at large scales, and only $J_{\ell\ell'}^\text{interp}$ is able to reliably recover the underlying signal.
However, like the full transfer matrix $J_{\ell\ell'}$ itself, $J_{\ell\ell'}^\text{interp}$ is ill-conditioned and care needs to be taken in solving the linear system.
$\Lambda$CDM-like $TT$, $EE$, and $BB$ spectra are easily recovered using a TSVD approach, but a dust foreground-like spectrum requires more sophisticated algorithms like the BiCGSTAB method to remain numerically stable.

Nonetheless, because \Spider employs a map-based method of foreground removal, we expect the remaining signal to deviate, at most, only slightly from $\Lambda$CDM.
As long as this is the case, the error from binning remains small (\hyperref[sec:appendix]{Appendix}).
We therefore chose to use the binned transfer matrix $J_{bb'}$ in the NSI pipeline for \Spider's main result \citep{bmode_paper}, as it reliably recovers a $\Lambda$CDM-like spectrum (Figure \ref{fig:xferratios}) and is easily invertible.
However, we caution that $J_{bb'}$ should only be pursued when the binning scheme used in analysis has been established, as its input bins must match the output bins; any changes to the binning scheme must be accompanied by a re-computation of $J_{bb'}$.
On the other hand, $J_{\ell\ell'}^\text{interp}$ is only as accurate as the interpolation allows, but is attractive for the capacity to increase its resolution as desired.
Additionally, $J_{\ell\ell'}^\text{interp}$ may be most advantageous for large angular scale foreground analyses, where it performs significantly better than $J_{bb'}$.

Of course, there is no need to adhere strictly to one technique.
If computational resources are limited, one may employ a hybrid approach and compute the transfer matrix only in regimes dominated by mode--mode coupling, \emph{i.e.}, where the transfer matrix has significant off-diagonal components.
For example, one may choose to commit to fully computing every column of the lowest bin, where the discrepancy between the input and output spectra is greatest (see Figure \ref{fig:xferratios}), and use a less expensive strategy elsewhere.
We also leave open the exploration of different interpolation schemes, as the implementation described in this paper has not been found to significantly affect \Spider's $B$-mode results.
Alternatively, the simple structure of the transfer matrix in Fourier space also suggests, in place of interpolation, a model with a polynomial fit in magnitude and a linear fit in phase.
We note that the biggest bottleneck is often computational time for the production of re-observed maps; as long as this portion of the pipeline is fixed, downstream tasks such as masking and choice of spectrum estimator are trivial to change.

As CMB experiments become capable of more precise measurements, improving the accuracy of the signal recovery process becomes increasingly important.
The transfer matrix decouples the estimation of power spectra -- and, by extension, the cosmological parameters -- from assumptions about the shape of the underlying sky signal.
This is particularly useful when the cosmological signals are heavily obscured by foreground Galactic dust, and is thus relevant for upcoming CMB instruments searching for primordial $B$-modes \citep[\emph{e.g.},][]{CMBS4_CDR, SO_science_book, hazumi2020litebird, shaw_spie} or measuring the $E$-modes on the largest angular scales \citep{allison2015, watts2018projected, hazumi2020litebird}.
As foreground component separation becomes an increasingly important part of CMB analyses, future experiments can benefit from the development of new signal-independent analysis techniques to improve the power spectrum reconstruction.

\section*{Acknowledgments}
\Spider is supported in the U.S. by the National Aeronautics and Space Administration under grants NNX07AL64G, NNX12AE95G, NNX17AC55G, and 80NSSC21K1986 issued through the Science Mission Directorate and by the National Science Foundation through PLR-1043515.
Logistical support for the Antarctic deployment and operations is provided by the NSF through the U.S. Antarctic Program.
Support in Canada is provided by the Natural Sciences and Engineering Research Council and the Canadian Space Agency.
Support in Norway is provided by the Research Council of Norway.
Support in Sweden is provided by the Swedish Research Council through the Oskar Klein Centre (Contract No.\ 638-2013-8993) as well as a grant from the Swedish Research Council (dnr.\ 2019-93959) and a grant from the Swedish Space Agency (dnr.\ 139/17).
The Dunlap Institute is funded through an endowment established by the David Dunlap family and the University of Toronto.
The multiplexing readout electronics were developed with support from the Canada Foundation for Innovation and the British Columbia Knowledge Development Fund.
KF holds the Jeff \& Gail Kodosky Endowed Chair at UT Austin and is grateful for that support.
WCJ acknowledges the generous support of the David and Lucile Packard Foundation, which has been crucial to the success of the project.
CRC was supported by UKRI Consolidated Grants, ST/P000762/1,
ST/N000838/1, and ST/T000791/1.

Some of the results in this paper have been derived using the \texttt{HEALPix} package \citep{HealPix}.
The computations described in this paper were performed on four computing clusters: Hippo at the University of KwaZulu-Natal, Feynman at Princeton University, and the GPC and Niagara supercomputers at the SciNet HPC Consortium \citep{Scinet,Niagara}.
SciNet is funded by the Canada Foundation for Innovation under the auspices of Compute Canada, the Government of Ontario, Ontario Research Fund -- Research Excellence, and the University of Toronto.

The collaboration is grateful to the British Antarctic Survey, particularly Sam Burrell, and to the Alfred Wegener Institute and the crew of R.V.  {\it Polarstern} for invaluable assistance with the recovery of the data and payload after the 2015 flight.
Brendan Crill and Tom Montroy made significant contributions to \Spider's development.  Paul Steinhardt provided very helpful comments regarding the status of early universe models.
This project, like so many others that he founded and supported, owes much to the vision and leadership of the late Professor Andrew E. Lange.

\bibliographystyle{aasjournal}
\bibliography{references_publishedversion}

\appendix
\section{Relation between Bin--Bin and Multipole--Multipole Transfer Matrices}
\label{sec:appendix}

The power spectrum binning procedure can be described by an operator $P_{b\ell}$ that determines the relative weighting of each multipole $\ell$ within a bin $b$:
\begin{align}
    \label{eq:cb_def}
    C_b = \sum\limits_{\ell} P_{b\ell} C_{\ell}.
\end{align}
Likewise, we can bin the pseudo-power spectrum and write
\begin{align}
    \label{eq:pseudo_cb}
    \tilde{C}_b = \sum\limits_{\ell} P_{b\ell} \tilde{C}_{\ell} = \sum\limits_{\ell} P_{b\ell} \sum\limits_{\ell'} J_{\ell\ell'} C_{\ell'}
\end{align}
as a consequence of Equation \ref{eq:jll_def}. Combining Equations \ref{eq:jbb_def} and \ref{eq:pseudo_cb},
\begin{align}
    \label{eq:jbb_and_jll}
    \sum\limits_{b'} J_{bb'} \sum\limits_{\ell'} P_{b'\ell'} C_{\ell'} = \sum\limits_{\ell} P_{b\ell} \sum\limits_{\ell'} J_{\ell\ell'} C_{\ell'}.
\end{align}
From this equation, we can identify the binned transfer matrix $J_{bb'}$ as taking the form
\begin{align}
    \label{eq:jbb_final}
    J_{bb'} = \frac{\sum\limits_{\ell} P_{b\ell} \sum\limits_{\ell' \in b'} J_{\ell\ell'} C_{\ell'}}{\sum\limits_{\ell'} P_{b'\ell'} C_{\ell'}}.
\end{align}
Note that the inner sum in the numerator is confined to the input bin $b'$ -- recall each column of $J_{bb'}$ is the output of passing a single input bin through the re-observation pipeline (in analogy with the unit $\delta$-functions used to construct $J_{\ell\ell'}$).
Taking the sum over all input bins $b'$, as on the left-hand side of Equation \ref{eq:jbb_and_jll}, recovers the sum over all input multipoles $\ell'$, as on the right-hand side.
For \textsc{Spider}, we employ a uniform weighting,
\begin{align}
    \label{eq:pbl}
    P_{b\ell} = \begin{cases} \frac{1}{\Delta\ell} & \text{if } \ell \in b, \\ 0 & \text{otherwise}, \end{cases}
\end{align}
which reduces Equation~\ref{eq:jbb_final} to
\begin{equation}
    \label{eq:jbb_special}
    J_{bb'} = \sum\limits_{\ell \in b} \left( \frac{\sum\limits_{\ell' \in b'} J_{\ell\ell'} C_{\ell'}}{\sum\limits_{\ell' \in b'} C_{\ell'}} \right).
\end{equation}
The quantity in parentheses in Equation \ref{eq:jbb_special} can be taken to be $J_{\ell b'}$, the proportion of the power in input bin $b'$ that contributes to the power in output multipole $\ell$.
It is equal to the weighted average of all of the $J_{\ell\ell'}$ elements that lie within input bin $b'$, with the weights given by some input model spectrum $C_{\ell}$.
This highlights an important point: while $J_{\ell\ell'}$ is a function of the data analysis pipeline only, $J_{bb'}$ is additionally dependent on the input spectrum used to create it.

We can understand this point by noting how $J_{\ell\ell'}$ and $J_{bb'}$ are each applied to a target power spectrum $c_\ell$:
\begin{subequations}
\begin{enumerate}
    \item First apply $J_{\ell\ell'}$, then bin the result:
        \begin{align}
            \tilde{c}_b = \frac{1}{\Delta\ell} \sum\limits_{\ell\in b} \sum\limits_{\ell'} J_{\ell\ell'} c_{\ell'}
            \label{eq:applyjll}
        \end{align}
    \item First bin $c_\ell$, then apply $J_{bb'}$:
        \begin{align}
            \hat{c}_b = \sum\limits_{b'} \left( \frac{\sum\limits_{\ell\in b} \sum\limits_{\ell'\in b'} J_{\ell\ell'} C_{\ell'}}{\sum\limits_{\ell'\in b'} C_{\ell'}} \right) \left( \frac{1}{\Delta\ell} \sum\limits_{\ell'\in b'} c_{\ell'} \right)
            \label{eq:applyjbb}
        \end{align}
\end{enumerate}
\end{subequations}
If we let $\delta_\ell$ represent the deviation of $c_\ell$ from $C_\ell$ (\emph{i.e.}, $c_\ell = C_\ell + \delta_\ell$; not to be confused with a $\delta$-function centred at $\ell$), then Equations \ref{eq:applyjll} and \ref{eq:applyjbb} can be understood intuitively:
whereas $\tilde{c}_b$ represents the true amount of filtered power in bin $b$, $\hat{c}_b$ is an estimate that differs from the true value by
\begin{align}
    \delta_b = \tilde{c}_b - \hat{c}_b = \frac{1}{\Delta\ell} \left( \sum\limits_{\ell\in b} \sum\limits_{\ell'} J_{\ell\ell'} \delta_{\ell'} - \sum\limits_{b'} J_{bb'} \sum\limits_{\ell'\in b'} \delta_{\ell'} \right).
    \label{eq:error}
\end{align}
In other words, the error from using $J_{bb'}$ is the difference between the average of transformed excess power in each multipole $\ell$ and transform of average excess power in each bin $b$.
Note that as $\delta_\ell \rightarrow 0$, the deviation $\delta_b \rightarrow 0$ as expected, \emph{i.e.}, $J_{bb'}$ returns the same result as $J_{\ell\ell'}$ when applied to the input spectrum $C_\ell$ used to create it.
Consequently, the choice of input spectrum should be as close as possible to the target spectrum $c_\ell$ to be applied, such that $\delta_b$ is minimized.

\end{document}